\documentclass[12pt]{article}  
\usepackage{amsmath,amssymb,amsfonts}
\usepackage[dvipdfm]{graphicx}
\usepackage{longtable}

\def\mylist{\begin{list}{}{\setlength{\leftmargin}{0.5in}
               \setlength{\listparindent}{-0.5in}
               \setlength{\itemindent}{\listparindent}}}

\newcommand{\nD}[1]{\not D}
\newcommand{\Tr}{{\rm Tr}}

\begin{document}

\begin{titlepage}

\title{Examples of global symmetry enhancement by monopole operators}
\author{Denis Bashkirov\\ {\it California Institute of Technology}\\ E-mail: bashkirov@theory.caltech.edu}

\maketitle

\abstract{We consider examples of global symmetry enhancement by monopole operators in three dimensional ${\cal N}=4$ gauge theories. These examples include unitary overbalanced quivers, quivers with non-simply laced gauge groups and nonlinear quivers.}

\end{titlepage}

\section{Introduction.}

  In this paper we continue the investigation of hidden symmetries in supersymmetric gauge theories in three dimensions using the method developed in \cite{GW},\cite{BK},\cite{Kim}.
The emphasis is now placed upon global symmetries in the infrared limit of ${\cal N}=4$ theories whose currents do not lie in the same ${\cal N}=4$ supermultiplet as the stress tensor. The lowest components of the global symmetries multiplets are scalars with conformal dimension 1 and in the adjoint representation of the corresponding group of global symmetries.

The ${\cal N}=4$ theories have microscopic $SO(4)_R\simeq SU(2)_R\times SU(2)_N$ R-symmetry. Because the R-symmetry group of a superconformal ${\cal N}=4$ theory to which the microscopic theory flows in the IR is $SO(4)$, one is tempting to assume the equality between the microscopic and the superconformal R-groups. This is known to be a wrong assumption in general, as there may appear accidental symmetries in the infrared whose currents are not conserved along the full RG flow from the ultraviolet. Luckily, in a large class of models the IR superconformal R-symmetry group is the microscopic one. Although, given a UV theory, it is not known how to prove this statement, there is a necessary condition for $R_{UV}=R_{IR}$ to hold which is easy to check: if, with respect to any subgroup $U(1)\subset SO(4)_R^{UV}$ there is a chiral operator with nonpositive R-charge, the IR R-symmetry is not the microscopic one. This condition is a simple consequence of unitarity in the IR\footnote{See \cite{GW} for discussion of this point and some examples.}. Moreover, in all known cases where the infrared R-symmetry is not the UV R-symmetry this manifests itself by appearence of chiral operators with R-charges violating unitarity. Thus it seems reasonable to assume the condition to be suffucient as well.

\section{Models.}

  We consider three-dimensional ${\cal N}=4$ supersymmetric gauge theories with a compact group $G$.  The fields form two ${\cal N}=4$ multiplets: vector multiplet ${\cal V}$ consisting of a vector ${\cal N}=2$  $V$ and a chiral ${\cal N}=2$ $\Phi$ multiplets in the adjoint representation of the gauge group $G$, and a hypermultiplet $\cal{H}$ consisting of two chiral ${\cal N}=2$ multiplets $Q$ and $\tilde Q$ in fundamental and anti-fundamental representations of $G$, respectively. Each chiral multiplet contains a complex scalar and a complex spinor (two Majorana spinors). The vector multiplet has a gauge field, a real scalar (dimensional reduction from a gauge field in 4d) and a complex spinor.

\section{A brief review of monopole operators.}

Here we recap the basic facts about monopole operators in three dimensional gauge theories \cite{BKW}.

  By definition, a hidden symmetry is generated by a conserved current whose existence does not follow from any symmetry of an action. The simplest example of such a symmetry corresponds to a topological conserved current which exists in any 3d gauge theory whose gauge group contains a $U(1)$ factor:
\begin{align}
J^\mu=\frac{1}{2\pi}\epsilon^{\mu\nu\lambda}\Tr\, F_{\nu\lambda}
\end{align}
There may be more complicated hidden symmetries whose conserved currents are monopole operators, i.e. disorder operators defined by the condition that the gauge field has a Dirac monopole singularity at the insertion point. More concretely, in a $U(N)$ gauge theory the singularity corresponding to a monopole operator must have the form
\begin{align}
A^{N,S}(\vec{r})=\frac{H}{2}(\pm1-\cos\,\theta)d\phi 
\end{align}
for the north and south charts, correspondingly. In this formula $H=diag(n_1,n_2,\ldots,n_N)$ and integers $n_1\ge n_2\ge\ldots \ge n_N$ are called magnetic charges (or GNO charges \cite{GNO}). If we require the monopole operator to preserve some supersymmetry (such operators may be called BPS operators),  matter fields must also be singular, in such a way  that BPS equations are satisfied in the neighborhood of the insertion point.

In special curcimstances it is possible to determine the spectrum of chiral monopole  operators with low values of conformal dimension. Namely, if we have a superconformal theory we can implement the radial quantization to obtain a supersymmetric theory on ${\mathbb R}\times S^2$ whose states are in one-to-one correspondence with local operators of the original theory on ${\mathbb R^3}$. The quantum numbers match on both sides of the correspondence with energies of the states being equal to conformal dimensions of the corresponding operators. For ${\cal N}=4$ gauge theories with vanishing anomalous dimensions of operators it is possible to continuously deform the theory on the sphere in a controlled supersymmetric way to a free theory of fields in a fixed spherically symmetric background determined by the Dirac monopole singularity at the insertion point (\cite{Kim}). It is then possible to find the spectrum of chiral monopole operators with lowest values of conformal dimensions (\cite{BK}). In the absence of Chern-Simons couplings the lowest energy states in the radial quantization are bare monopoles, that is, "vacuum" states in sectors determined by magnetic charges which are not excited with fields modes. 

\section{Review of the method.}

In their paper \cite{IS} Intriligator and Seiberg suggested a dual description of the infrared limit of a class of three dimensional ${\cal N}=4$ quiver gauge theories. A particular prediction of this correspondence was presence of hidden global symmetries in the quiver theories. It was realized long ago that conserved currents that span the cartan subalgebra are simply the topological currents of each of the U(1) factors of the compact gauge group (see \cite{BKW}). The realization of the rest of the global currents was claimed to be by means of monopole operators \cite{BKW}. This claim was verified in paper \cite{GW} whose authors showed that all the neccesary currents are monopole operators by finding all chiral scalars of conformal dimension 1 and making sure that their topological charges are exactly the ones appropriate for the roots of the global symmetry groups. This means that the conserved currents produced from them taken together with the topological currents form the required Lie algebra. 

 Moreover, authors of \cite{GW} showed that any quiver whose nodes are 1) unitary groups, 2) balanced nodes is necessary one of the $ADE$ quivers.
A node corresponding to gauge group $U(n_c)$ and $n_f$ fundamental hypermultiplets\footnote{Here $U(n_c)$ is treated in isolation from the other nodes in the sense that all bifundamentals and fundamentals themselves are included in $n_f$. For example, a bifundamental hyper of $U(n_c)\times U(N)$ is considered as $n_f=N$ fundamental hypers.} is called balanced if $n_f=2n_c$. They also conjectured that a quiver each node of which is good ($n_f\ge 2n_c$) is good in whole, that is, all chiral monopole operators have $E\ge 1$ and the corresponding theory flows to the standard IR limit (the $R$-symmetry is the microscopic one) with monopole symmetries being products of $ADE$ groups and $U(1)$.   

In this paper we provide examples of such quiver theories and find their monopole symmetries.

In order to establish notations and illustrate the procedure we review the cases of $D_4$ and $D_5$ quivers that will also serve as starting points for deformed quivers with new global symmeties considered later in the paper.

\subsection{$D_4$ and $D_5$ quivers.}

Let us start with the smallest quiver of D-type corresponding to the extended Dynkin diagram $D_4$. This diagram represents the Lie algebra $so(8)$.
It translates to a quantum field theory as follows.
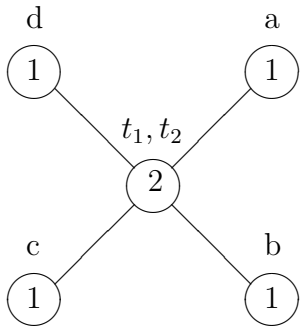
\begin{figure}
\begin{center}
\begin{picture}(180,140)(-100,-50)
\put(8,22){\line(-1,1){30}}
\put(8,8){\line(-1,-1){30}}
\put(-30,58){\circle{20}}
\put(-33,55){1}
\put(-33,75){d}
\put(-30,-28){\circle{20}}
\put(-33,-31){1}
\put(-33,-11){c}
\put(15,15){\circle{20}}
\put(13,13){2}
\put(3,33){$t_1,t_2$}
\put(22,22){\line(1,1){30}}
\put(22,8){\line(1,-1){30}}
\put(60,58){\circle{20}}
\put(57,55){1}
\put(57,75){a}
\put(60,-28){\circle{20}}
\put(57,-31){1}
\put(57,-11){b}
\end{picture}
\caption{\small $D_4$ quiver. Letters stand for magnetic fluxes.}\label{D1}
\end{center}
\end{figure}

The central node with index $2$ denotes the gauge group $U(2)$, the other four nodes are U(1) gauge groups and the edges are bifundamental hypermultiplets. Because there are no fundamental hypers there is a decoupled U(1) gauge subgroup which is manifested in the invariance of the energy of bare chiral monopoles (\cite{GW},\cite{BK},\cite{Kim})

\begin{align}
& E=-|t_1-t_2|+\nonumber\\
& \frac{1}{2}(|t_1-b|+|t_2-b|+|t_1-c|+|t_2-c|+|t_1-d|+|t_2-d|+|t_1-a|+|t_2-a|)
\end{align}

under shifts by equal fluxes $\{t_1,t_2;b,c,d,a\}\to\{t_1+m,t_2+m;b+m,c+m,d+m,a+m\}$. To deal with this redundacy we fix 'the gauge' by setting $a=0$. It is important that no nonzero flux distributions give non-positive energy (after gauge fixing only the zero fluxes give zero energy). This means that the microscopic R-symmetry can be the R-symmetry that enters the superconformal algebra in the infrared. Note that energy positivity for bare monopoles is nontrivial in this case because the vector multiplet gives negative contribution.

A calculation gives 24 bare monopole scalars with energy 1 corresponding to different magnetic (and topological!) charges  (see appendix A). In the basis $(h_1,h_2,h_3,h_4)$ where $\{t_1+t_2=h_2-h_4, b=h_3+h_4, c=h_3-h_4, d=h_1-h_2\}$ it is obvious that the scalars together with 4 nontopological chiral scalars $tr\phi$ (they are superpartners of four topological currents and are lowest components  from the chiral multiplets $tr\Phi$) are in adjoint representation of ${so}(8)$.  This leads to 28 conserved currents forming Lie algebra ${so}(8)$.

A similar analysis can be performed for the quiver diagram $D_5$. It is shown in appendix A that this leads to a global symmetry group $SO(10)$ with its currents being monopole operators.

Note that we did not prove the absence of nonzero fluxes leading to nonpositive energies. The necessary condition $n_f\ge 2n_c-1$ for each node is obviously satisfied. Moreover the stronger condition $n_f=2n_c$ holds. The condition $n_f\ge 2n_c-1$ is necessary because if it is not satisfied one gets a bare monopole with nonpositive energy which is magnetically charged under the corresponding gauge subgroup and magnetically neutral under all the rest factors in the full gauge group. Authors of \cite{GW} showed that $D_n$ models have no bare monopoles with nonpositive energy.
 
\subsection{$E_{6,7,8}$-type quivers.}

The gauge group and the field content can be read off from figures B1, B2 and B3 in Appendix B.

 When we run the procedure from the previous subsection for $E_6$ we find lowest energy states being 72 bare monopole scalars of energy $E=1$ with magnetic charges just right to form an adjoint representation after completing them with 6 nonmagnetical chiral scalars -- traces $tr\phi$ of the chiral scalars which are the lowest components of the ${\cal N}=2$ chiral multiplets $\Phi$. Acting twice with supercharges on the 72 scalars we get conserved currents which correspond to roots of the global group $E_6$. The six independent topological charges correspond to the cartan operators.

For the $E_7$ quiver theory we find 126 bare monopoles with energy $E=1$ and appropriate topological charges\footnote{For all theories considered in this section to each set of topological charges there corresponds a unique set of magnetic charges.} leading to the existence of $E_7$ group of symmetries realized by monopole operators.

The situation for the $E_8$ quiver theory is similar: 240 bare monopoles with energy $E=1$ give rise to the $E_8$ symmetry of the theory.

\section{Engineering nonlinear quivers.}

  It turns out to be difficult if possible at all to construct quiver theories with hidden non ADE-type groups of global symmetries realized by monopole operators. More precisely, it is possible to build $Sp(N)$ with the symmetry currents lying in a free sector of the IR theory. Examples of such theories of linear-quiver type were given in \cite{GW}. In the next section we provide some examples of nonlinear-quiver theories with free symplectic symmetry group. However, it is very easy to build a large class of theories whose symmetry groups contain nonfree factors of A-D-E type.

Indeed, given an A-D-E-type theory consider connecting arbitrary theory to some nodes of the original quiver. This means that we take two theories A and B that contain gauge subgroups $G_A$ and $G_B$, correspondingly, as factors and add a hypermultiplet in a (nontrivial) representation of $G_A\times G_B$. If the two original theories had monopole symmetries $S_A$ and $S_B$ then the engineered theory will have at least a subgroup of $S_A\times S_B$ corresponding to zero fluxes for $G_A\times G_B$. Note that the engineered theory $A\times B$ is always good if $A$ and $B$ are good theories separately. This is because the expression for the energy of bare monopoles in the engineered theory is a sum of those in the original theories and a positive contribution from the new hypermultiplet. Similarly, ugly theories produce an ugly or a good theory. Ugly means that the minimal non-zero energy of chiral operators is 1/2, that is, they are free. 

Let us consider several examples illustrating this construction taking the $D_n$ quiver theory as an original theory. 
\bigskip

{\it Example 1.}

\begin{figure}
\begin{center}
\begin{picture}(140,95)(-60,-35)
\put(8,22){\line(-1,1){30}}
\put(8,8){\line(-1,-1){30}}
\put(-30,58){\circle{20}}
\put(-33,55){1}
\put(-53,55){d}
\put(-30,-28){\circle{20}}
\put(-33,-31){1}
\put(-53,-31){c}
\put(-30,25){\line(0,1){23}}
\put(-30,15){\circle{20}}
\put(-33,13){1}
\put(-53,13){f}
\put(-30,-18){\line(0,1){23}}
\put(15,15){\circle{20}}
\put(13,13){2}
\put(22,22){\line(1,1){30}}
\put(22,8){\line(1,-1){30}}
\put(60,58){\circle{20}}
\put(57,55){1}
\put(77,55){a}
\put(60,-28){\circle{20}}
\put(57,-31){1}
\put(77,-31){b}
\end{picture}
\end{center}
\caption{\small {\it Example 1} quiver.}\label{D2}
\end{figure}
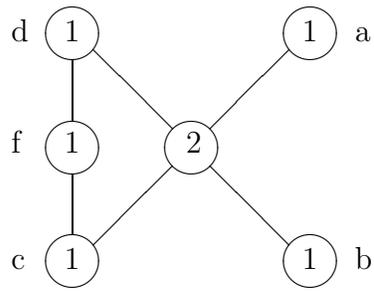

Consider the $D_4$ quiver as an $G_A$ theory and an $U(1)$ theory with two fundamental hypermultiplets as an $G_B$ theory and modify them to the $D_4\times U(1)$ gauge theory as in Figure \ref{D2}. This new theory has $SU(4)\times SU(2)\times U(1)$  as its symmetry group in the monopole sector. The first factor is inhereted from the original $D_4$ theory (it was possible for the new $U(1)$ factor to have such magnetic flux so that not to excite\footnote{An excited edge corresponds to a hypermultiplet that gives a nontrivial contribution to the energy of a bare monopole.} any new edge compared with the fluxes distribution producing the $SU(4)$ symmetry subgroup in $D_4$ theory) while the $SU(2)$ corresponds to putting one unit (and minus the unit) of flux for the $U(1)$ gauge factor while leaving all the rest fluxes zero. The last $U(1)$ factor is just one of the five topological charges under which none of the bare monopoles carries a charge. The whole symmetry group is nonfree.\footnote{Quivers of this type appeared in \cite{LRvU}.}

\bigskip
{\it Example 2.}

\begin{figure}
\begin{center}
\begin{picture}(140,95)(-60,-35)
\put(8,22){\line(-1,1){30}}
\put(8,8){\line(-1,-1){30}}
\put(-30,58){\circle{20}}
\put(-33,55){1}
\put(-53,55){d}
\put(-30,-28){\circle{20}}
\put(-33,-31){1}
\put(-53,-31){c}
\put(-30,-18){\line(0,1){8}}
\put(-30,29){\circle{20}}
\put(-33,26){1}
\put(-30,10){\line(0,1){9}}
\put(-30,0){\circle{20}}
\put(-30,39){\line(0,1){9}}
\put(-33,-3){1}
\put(15,15){\circle{20}}
\put(13,13){2}
\put(22,22){\line(1,1){30}}
\put(22,8){\line(1,-1){30}}
\put(60,58){\circle{20}}
\put(57,55){1}
\put(77,55){a}
\put(60,-28){\circle{20}}
\put(57,-31){1}
\put(77,-31){b}
%\put(-25,-50){\small Figure \ref{D3}. {\it Example 2} quiver.}
\end{picture}
\end{center}
\caption{{\it Example 2} quiver.}\label{D3}
\end{figure}
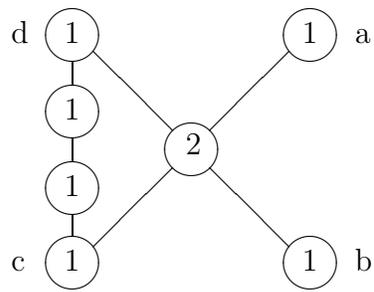

We start with a $D_4$ quiver again and add two more $U(1)$ factor and bifundamental hypermultiplets as in Fig.\ref{D3}. This gives an $SU(4)\times SU(3)\times U(1)$ nonfree monopole symmetry. 

\bigskip
{\it Example 3.}

\begin{figure}
\begin{center}
\begin{picture}(160,95)(-40,-35)
\put(8,22){\line(-1,1){30}}
\put(8,8){\line(-1,-1){30}}
\put(-30,58){\circle{20}}
\put(-33,55){1}
\put(-30,-28){\circle{20}}
\put(-33,-31){1}
\put(-30,-18){\line(0,1){23}}
\put(-30,15){\circle{20}}
\put(-33,13){1}
\put(-30,25){\line(0,1){23}}
\put(15,15){\circle{20}}
\put(13,13){2}
\put(25,15){\line(1,0){40}}
\put(75,15){\circle{20}}
\put(73,13){2}
\put(82,22){\line(1,1){30}}
\put(82,8){\line(1,-1){30}}
\put(120,58){\circle{20}}
\put(117,55){1}
\put(120,-28){\circle{20}}
\put(117,-31){1}
%\put(-15,-45){\small Figure \ref{qD4}. {\it Example 3} quiver.}
\end{picture}
\end{center}
\caption{\small {\it Example 3} quiver.}\label{qD4}
\end{figure}
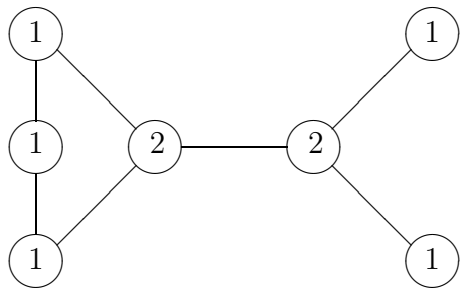

This time we take the $D_5$ quiver and add one $U(1)$ factor and bifundamental hypers as Fig. \ref{qD4}. The resulting monopole symmetry group is nonfree $SO(8)\times SU(2)\times U(1)$.

\bigskip
{\it Example 4.}

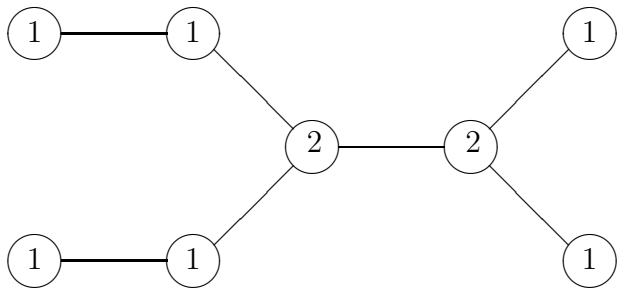
\begin{figure}
\begin{center}
\begin{picture}(220,105)(-100,-35)
\put(-80,58){\line(1,0){40}}
\put(-90,58){\circle{20}}
\put(-93,55){1}
\put(-80,-28){\line(1,0){40}}
\put(-90,-28){\circle{20}}
\put(-93,-31){1}
\put(8,22){\line(-1,1){30}}
\put(8,8){\line(-1,-1){30}}
\put(-30,58){\circle{20}}
\put(-33,55){1}
\put(-30,-28){\circle{20}}
\put(-33,-31){1}
\put(15,15){\circle{20}}
\put(13,13){2}
\put(25,15){\line(1,0){40}}
\put(75,15){\circle{20}}
\put(73,13){2}
\put(82,22){\line(1,1){30}}
\put(82,8){\line(1,-1){30}}
\put(120,58){\circle{20}}
\put(117,55){1}
\put(120,-28){\circle{20}}
\put(117,-31){1}
%\put(-15,-50){{\small {\it Example 4} quiver.}}
\end{picture}
\end{center}
\caption{\small {\it Example 4} quiver.}\label{qD5}
\end{figure}

Take the $D_5$ quiver and add two $U(1)$ factors as in Fig.\ref{qD5}. This time the whole $SO(10)$ group is preserved and there appears an additional factor $SU(2)_{free}\times SU(2)_{free}$ in the monopole symmetry group $SO(10)\times SU(2)_{free}\times SU(2)_{free}$. This happens because any original distribution of fluxes can be embedded in the new quiver without changing the net energy by simply putting fluxes on the new factors so that no new edge is excited, that is, no new hypermultiplet contributes to the energy. Moreover, now we can put fluxes on the two new $U(1)$ gauge group factors and set the rest fluxes to zero. This produces the $SU(2)_{free}\times SU(2)_{free}$ free factor. The subscript  'free' is used to stress that the currents of the corresponding group (or, equivalently, $E=1$ scalars) are built from free fields.  A natural guess then is that the infrared limit of this theory is that of $X\times X\times D_5$ with $X$ being the theory of a free twisted hypermultiplet. The bare monopoles with $E=1/2$ correspond to the lowest component scalars in the free twisted hypermultiplets. This also is in accord with the argument in favor of a similar factorization on page 24 of \cite{GW}.

\section{Quiver theories with nonunitary gauge groups.}

So far we have considered only theories whose gauge groups are products of unitary groups. Let us analize non-simply laced gauge groups $SO(5)$ and $G_2$.

The formula for energies of bare monopoles is trivially generalized to arbitrary gauge groups $G$.

\begin{align}
E(H)=-\frac12\sum_{r}|r(H)|+\frac12\sum_{w}|w(H)|=-\sum_{r^+}|r^+(H)|+\frac12\sum_{w}|w(H)|\label{E}
\end{align}
In this formula $H$ is the cartan generator containing magnetic charges, that is $e^{iH\alpha}$ is the image of $e^{i\alpha}$ under an embedding $U(1)\hookrightarrow G$ that defines magnetic (GNO) charges; $r$ stands for roots, $r^+$ for positive roots and $w$ for weights of representations of all hypermultiplets. Magnetic weights, or cartan generators $H$, can take any values satifying the Dirac quantization condition

\begin{align}
w(H)\in Z\quad\hbox{for all weights $w$ of all representations present}.\label{Dirac}
\end{align}

\subsection{$G_2$ case.}

$G_2$ is a rank two simple Lie algebra of dimension 14. The root space is a two dimensional vector space ${\mathbb R}^2$, in which positive simple roots can be taken as

\begin{align}
\alpha=(1,0),\quad \beta=(-\frac32,\frac{\sqrt{3}}{2}).
\end{align}

Other positive roots are

\begin{align}
r_3=\alpha+\beta,\quad r_4=2\alpha+\beta,\quad r_5=3\alpha+\beta,\quad r_6=3\alpha+2\beta.
\end{align}

The cartan algebra is a two dimensional vector space dual to ${\mathbb R}^2$ which can be identified with it by means of the standard metric. Let us chose a basis $\{H_1,H_2\}$ in it dual to $\{\alpha,\beta\}$ and write an arbitrary cartan as $H=n_1H_1+n_2H_2$ where $(n_1,n_2)$ are a priori real numbers. Then the contribution of the vector multiplet to the energy evaluated on such a cartan
is
\begin{align}
& E_v(H)=-(|\alpha(H)|+|\beta(H)|+|r_3(H)|+|r_4(H)|+|r_5(H)|+|r_6(H)|)=\nonumber\\
& -(|n_1|+|n_2|+|2n_1+n_2|+|n_1+n_2|+|3n_1+n_2|+|3n_1+2n_2|).
\end{align}

Comparing this expression with (\ref{E}) and (\ref{Dirac}) we conclude that $n_1$ and $n_2$ must be integral.
We consider three examples with a fundamental hypermultiplets. $G_2$ has two fundamental representations with highest weights $2\alpha+\beta$ and $3\alpha+2\beta$. We focus on the former. It has dimension 7 and weights

\begin{align}
& w_1=\alpha,\quad w_2=2\alpha+\beta,\quad w_3=\alpha+\beta,\nonumber\\
& w_4=-w_1,\quad w_5=-w_2,\quad w_6=-w_6,\quad w_7=0.
\end{align}

{\it Example 5.}

This is a theory with gauge group $G_2\times U(4)^2$ corresponding to the quiver with a bifundamental hyper of $G_2\times U(4)$ for both $U(4)$ factors\footnote{Note that $n_f=2n_c-1$ for both $U(n_c)$ factors.}. The symmetry group is $Sp(2)_{free}\times U(1)^2$. The corresponding monopole scalars with energy $E=1$ are in the Table\ref{G2}.

\begin{table}
\begin{tabular}{|l|l|l|}
\hline
$(m_1m_2m_3m_4,n_1n_2n_3n_4,p_1p_2)$ & $(m_1m_2m_3m_4,n_1n_2n_3n_4,p_1p_2)$ & $(m_1m_2m_3m_4,n_1n_2n_3n_4,p_1p_2)$\\
\hline
\hline
(-1000,-1000,00) & (-2000,0000,00) & (1-100,0000,00)\\
(-1000,1000,00) & (2000,0000,00) & \\
(1000,-1000,00) & (0000,-2000,00) & (0000,1-100,00)\\
(1000,1000,00) & (0000,2000,00) & \\
\hline 
\end{tabular}
\caption{{\it Example 5.} Bare monopole states with energy $E=1$.}\label{G2}
\end{table}

In addition there are four monopole operators with energy $E=1/2$ (Table\ref{G3}).

\begin{table}
\begin{tabular}{|l|}
\hline
$(m_1m_2m_3m_4,n_1n_2n_3n_4,p_1p_2)$ \\
\hline
\hline
 (-1000,0000,00)\\
 (1000,0000,00)\\
 (0000,-1000,00)\\
 (0000,1000,00)\\
\hline 
\end{tabular}
\caption{{\it Example 5.} Bare monopole states with energy $E=1/2$.}\label{G3}
\end{table}

On ${\mathbb R}^3$ they correspond to free chiral operators. The eight states in the first two columns of Table\ref{G2} and four states in Table\ref{G3} are naturally reproduced in the theory $X^2$ which is the product of two free twisted hypermultiplets. In fact, by the argument of \cite{GW} on page 24 the whole theory is equivalent to a product $X\times X\times{\cal H}$ where ${\cal H}$ is a good theory with gauge group $G_2\times U(3)^2$ and bifundamental hypermultiplets for the two pairs $G_2\times U(3)$. The two topologically neutral states in the third column of Table\ref{G2} correspond to two scalars $tr\Phi$ of the two gauge $U(3)$. They are the lowest components of the multiplets containing topological currents for the two $U(3)$ gauge factors. If we add a fundamental hypermultiplet for each $U(4)$ we obtain the nonfree symmetry group $SU(2)_{nonfree}\times SU(2)_{nonfree}$. One can generalize this example by considering the gauge group $G_2\times U(4)^N$ with a bifundamental hypermutiplet for each pair $G_2\times U(4)$ for arbitrary natural number $N$. This gives the monopole symmetry group $Sp(N)_{free}\times U(1)^N$.

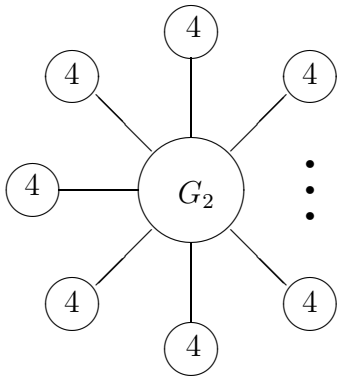
\begin{figure}
\begin{center}
\begin{picture}(120,160)(-50,-55)
\put(0,30){\line(-1,1){21}}
\put(0,0){\line(-1,-1){21}}
\put(15,75){\circle{20}}
\put(13,73){4}
\put(-30,58){\circle{20}}
\put(-33,55){4}
\put(15,35){\line(0,1){30}}
\put(-30,-28){\circle{20}}
\put(-33,-31){4}
\put(15,15){\circle{40}}
\put(10,10){$G_2$}
\put(30,30){\line(1,1){21}}
\put(60,58){\circle{20}}
\put(57,55){4}
\put(30,0){\line(1,-1){21}}
\put(-45,15){\circle{20}}
\put(-48,13){4}
\put(-35,15){\line(1,0){30}}
\put(60,-28){\circle{20}}
\put(57,-31){4}
\put(15,-45){\circle{20}}
\put(13,-48){4}
\put(15,-35){\line(0,1){30}}
\multiput(60,5)(0,10){3}%
{\circle*{3}}
\end{picture}
\end{center}
\caption{\small {\it Example 5} quiver.
 All lines stand for representations ${\bf 7}\times{\bf 4}$ of $G_2\times U(4)$.}\label{g1}
\end{figure}

\bigskip

{\it Example 6.}

As a gauge group we take $G_2\times U(4)$ with a bifundamental hypermultiplet and two fundamentals of $G_2$ needed to exclude chiral monopole operators with nonpositive energies. We use gauge groups with $U(1)$ factors to get nonabelian monopole symmetries. Each $U(1)$ factor provides a conserved topological current whose charge serves as a cartan. Because magnetic charges are not produced from any conserved currents simple gauge groups can only produce abelian monopole symmetries. The global symmetry group\footnote{In all cases we mention a (global) symmetry group we refer to the part of the symmetry group generated by monopole operators.} is $SU(2)_{free}\times U(1)$. The $SU(2)_{free}$ factor is free in the sense that the currents are build from a doublet of free fields which are bare monopole operators with energy $E=1/2$. By the conformal algebra such chiral operators are free fields. These fields carry magnetic charges only with respect to the $U(4)$ factor. Alternatively, this theory can be described as the infrared limit of $X\times {\cal H}$ where $X$ is the theory of a free twisted hypermultiplet and ${\cal H}$ is the original theory with $U(4)$ gauge group replaced by $U(3)$. Adding a fundamental hyper of $U(4)$ eliminates the free doublet and reduces the symmetry group to $SU(2)$ which is now nonfree. This quiver describes a 'good theory' in the terminology of \cite{GW}.

We could also take one fundamental of $G_2$. This theory has the same symmetry group as the above.

\subsection{$SO(5)$ case.}

The rank of the group is two and the dimension is 10. The roots are

\begin{align}
& \alpha_1=(1,0),\quad \alpha_2=(0,1),\quad\alpha_3=(1,1),\quad\alpha_4=(1,-1),\nonumber\\
&\alpha_5=-\alpha_1,\quad\alpha_6=-\alpha_2,\quad\alpha_7=-\alpha_3,\quad\alpha_4=-\alpha_4,
\end{align}
where $\alpha_2$ and $\alpha_4$ are positive simple roots.
The basis of cartans $\{H_1,H_2\}$ are chosen to be dual to $\{\alpha_1,\alpha_2\}$. On a cartan $H=n_1H_1+n_2H_2$ the contribution of the vector multiplet to the energy of bare monopoles is
\begin{align}
E_v=-(|n1|+|n2|+|n1-n2|+|n1+n2|),
\end{align}
so the magnetic charges $n_1$ and $n_2$ are at least integer.

The weights of a fundamental representation ${\bf 5}$ are 
\begin{align}
& w_1=\alpha_1,\quad w_2=\alpha_2,\quad w_3=-\alpha_1,\quad w_4=-\alpha_2,\quad w_2=0.
\end{align}

\bigskip
{\it Example 7.}

The gauge group is $SO(5)\times U(3)$ with one hypermultiplet in the bifundamental representation and two hypermultiplets in representation ${\bf 5}\times{\bf 1}$. The symmetry group is $SU(2)_{free}\times U(1)^3$. The two bare monopoles in the adjoint of $SU(2)$ have magnetic charges
\begin{align}
(0,0;-2,0,0),\quad (0,0; 2,0,0)
\end{align}

The three bare monopoles corresponding to the three $U(1)^3$ currents are
\begin{align}
(-1,0;1,-1,0),\quad (0,0;1,-1,0),\quad (1,0;1,-1,0)
\end{align}
We see that although the three $U(1)^3$ currents do not carry topological charge, they are magnetically charged with respect to both gauge subgroups. This $U(1)^3$ symmetry is nonfree.

It is possible to give a discription of this theory in which the free part and the interacting part of the IR theory are factorized already in the UV Lagrangian \cite{GW}. This theory is a product $X\times {\cal H}$ of the free twisted hypermultiplet $X$ and theory ${\cal H}$ which is obtained from the original one by replacing $U(3)$ gauge factors with $U(2)$ factors. In this description one of the $U(1)$ currents is the topological current of $U(2)$ while the other two are magnetically charged with respect to $SO(5)$ currents.

\bigskip
{\it Example 8.}

The gauge group is $SO(5)\times U(3)^N$ with a bifundamental of $SO(5)\times U(3)$ for each $U(3)$. The symmetry group is $Sp(N)_{free}\times U(1)^{N}_{nonfree}$\footnote{For $N=2$ there is an additional $U(1)^2$ symmetry with currents magnetically charged under $SO(5)$ but topologically neutral.}. There are bare monopoles with energy $E=1/2$ with nozero magnetic charges for only one of the $U(3)$ factors and bare monopoles with energy $E=1$ with nonzero magnetic charges for any two of the $U(3)$ factors. As in previous examples we can give a description of this theory where the factorization of the free sector is manifest already in the UV. This theory is $X^N\times {\cal H}$ where ${\cal H}$ is obtained from the original theory by replacing all $U(3)$ gauge factors by $U(2)$ factors.

 In this example we meet a certain universality. Because for all $E=1/2$ and $E=1$ bare monopoles $SO(5)$ magnetic charges are zero we can reproduce all the scalars by putting any other group in the center of the quiver instead of $SO(5)$ as long as it has a five-dimensional representation and the resulting theory does not have any bare monopoles with nonpositive energy. For example, we can take $SU(2)\times U(3)^N$ with hypers in representations ${\bf 5}\times {\bf 3}$ for each $U(3)$.

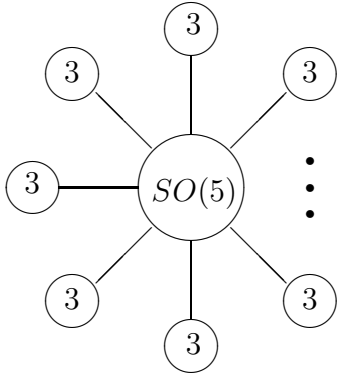
\begin{figure}
\begin{center}
\begin{picture}(120,130)(-50,-50)
\put(0,30){\line(-1,1){21}}
\put(0,0){\line(-1,-1){21}}
\put(15,75){\circle{20}}
\put(13,73){3}
\put(-30,58){\circle{20}}
\put(-33,55){3}
\put(15,35){\line(0,1){30}}
\put(-30,-28){\circle{20}}
\put(-33,-31){3}
\put(15,15){\circle{40}}
\put(0,10){$SO(5)$}
\put(30,30){\line(1,1){21}}
\put(60,58){\circle{20}}
\put(57,55){3}
\put(30,0){\line(1,-1){21}}
\put(-45,15){\circle{20}}
\put(-48,13){3}
\put(-35,15){\line(1,0){30}}
\put(60,-28){\circle{20}}
\put(57,-31){3}
\put(15,-45){\circle{20}}
\put(13,-48){3}
\put(15,-35){\line(0,1){30}}
\multiput(60,5)(0,10){3}%
{\circle*{3}}
\end{picture}
\end{center}
\caption{{\it Example 8} quiver.}\label{so}
\end{figure}

If we take all hypermultiplets in representation ${\bf 6}\times{\bf 3}$ of $SU(2)\times U(3)$ the monopole symmetry becomes nonfree $SU(2)\times SU(2)$.

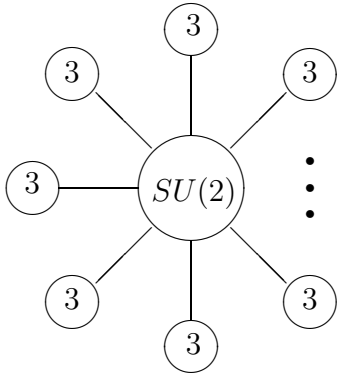
\begin{figure}
\begin{center}
\begin{picture}(120,130)(-50,-50)
\put(0,30){\line(-1,1){21}}
\put(0,0){\line(-1,-1){21}}
\put(15,75){\circle{20}}
\put(13,73){3}
\put(-30,58){\circle{20}}
\put(-33,55){3}
\put(15,35){\line(0,1){30}}
\put(-30,-28){\circle{20}}
\put(-33,-31){3}
\put(15,15){\circle{40}}
\put(0,10){$SU(2)$}
\put(30,30){\line(1,1){21}}
\put(60,58){\circle{20}}
\put(57,55){3}
\put(30,0){\line(1,-1){21}}
\put(-45,15){\circle{20}}
\put(-48,13){3}
\put(-35,15){\line(1,0){30}}
\put(60,-28){\circle{20}}
\put(57,-31){3}
\put(15,-45){\circle{20}}
\put(13,-48){3}
\put(15,-35){\line(0,1){30}}
\multiput(60,5)(0,10){3}%
{\circle*{3}}
\end{picture}
\end{center}
\caption{{\it Example 8} quiver. \small All lines stand for representations ${\bf 5}\times{\bf 3}$ of $SU(2)\times U(3)$.}\label{su}
\end{figure} 

\bigskip

{\it Example 9.}

The gauge group is $SO(5)\times U(2)$ but the theory is not of a quiver type because we take a hypermultiplet in the representation ${\bf 10}\times {\bf 4}$ and two hypers in the fundamental representation of $U(2)$. The symmetry group is $SU(2)_{nonfree}$. Again, the magnetic charges of $SO(5)$ are zero.\footnote{{\bf 10} is the adjoint representation of $SO(5)$ and {\bf 4} is the adjoint of $U(2)$.}

\subsection{Unitary quivers.}

\bigskip

{\it Example 10.}

Guided by the principle to take $n_f\ge2n_c-1$ for unitary quivers we can build the quiver theory depicted in Fig.B4 in Appendix B.
It turns out to have a nonfree $SO(14)\times U(1)$ monopole symmetry. The $U(1)$ factor is just one of the eight topological charges (one topological charge corresponds to the decoupled $U(1)_{diag}$ which gives eight topological charges) under which no $E=1$ bare monopoles are charged.

\bigskip

{\it Example 11.}
\bigskip 

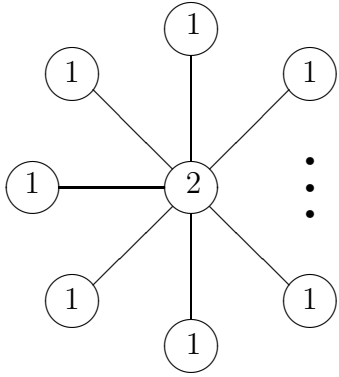
\begin{figure}
\begin{center}
\begin{picture}(120,140)(-50,-60)
\put(8,22){\line(-1,1){30}}
\put(8,8){\line(-1,-1){30}}
\put(15,75){\circle{20}}
\put(13,73){1}
\put(-30,58){\circle{20}}
\put(-33,55){1}
\put(15,25){\line(0,1){40}}
\put(-30,-28){\circle{20}}
\put(-33,-31){1}
\put(15,15){\circle{20}}
\put(13,13){2}
\put(22,22){\line(1,1){30}}
\put(60,58){\circle{20}}
\put(57,55){1}
\put(22,8){\line(1,-1){30}}
\put(-45,15){\circle{20}}
\put(-48,13){1}
\put(-35,15){\line(1,0){40}}
\put(60,-28){\circle{20}}
\put(57,-31){1}
\put(15,-45){\circle{20}}
\put(13,-48){1}
\put(15,-35){\line(0,1){40}}
\multiput(60,5)(0,10){3}%
{\circle*{3}}
\end{picture}
\end{center}
\caption{\small {\it Example 11} quiver.}\label{cham}
\end{figure}

Consider a quiver with gauge group $U(2)\times U(1)^N$ with bifundamental hypermultiplets for each subgroup $U(2)\times U(1)$ as in Fig.\ref{cham}. This theory has a nonfree symmetry $SU(2)^N$ except in the case $N=4$ which corresponds to quiver $D_4$ and enhanced symmetry $SO(8)$. The enhancement happens because for this particular value of $N$ the central node can have nonzero magnetic flux without spoiling condition $E=1$.

\section{Appendix A}

\subsection{D-type quivers.}

For $D_4$-quiver theory with bare monopole energy

\begin{align}
& E=-|t_1-t_2|+\nonumber\\
& \frac{1}{2}(|t_1-b|+|t_2-b|+|t_1-c|+|t_2-c|+|t_1-d|+|t_2-d|+|t_1-a|+|t_2-a|)
\end{align}

fixing the shift symmetry by setting $s=0$ we get 24 E=1 scalars. They are devided in two equal parts. One obtained from the other by flipping signs of magnetic charges. This is obvious because the expression for the energy is invariant under flipping the signs as well as our 'gauge fixing' condition.  The topological charges for one of the parts are given in Table \ref{D4}.

\begin{table}
\begin{tabular}{|l|l|l|l|l|l|l|}
\hline
tbcd & tbcd & tbcd & tbcd & tbcd & tbcd\\
\hline
\hline
0001 & 0010 & 1000 & 1001 & 1010 & 1011\\
0100 & 1100 & 1101 & 1110 & 1111 & 2111\\
\hline 
\end{tabular}
\caption{Positive topological charges of $D_4$ theory.}\label{D4}
\end{table}

For a new basis $(h_1,h_2,h_3,h_4)$ where

\begin{align}
c=h_3-h_4,\quad b=h_3+h_4,\quad t=h_2-h_4,\quad d=h_1-h_2
\end{align}

they are in Table \ref{D'4}.

\begin{table}
\begin{tabular}{|l|l|l|l|l|l|l|}
\hline
$h_1h_2h_3h_4$ & $h_1h_2h_3h_4$ & $h_1h_2h_3h_4$ & $h_1h_2h_3h_4$ & $h_1h_2h_3h_4$ & $h_1h_2h_3h_4$\\
\hline
\hline
1-100 & 001-1 & 01-10 & 0011 & 10-10 & 010-1\\
100-1 & 0101 & 1001 & 0110 & 1010 & 1100\\
\hline 
\end{tabular}
\caption{Positive topological charges of $D_4$ in new basis.}\label{D'4}
\end{table}

This is a set of positive roots of $SO(8)$ \cite{Simon}.

{\it $D_5$-type quiver theory.}

\begin{table}
\begin{tabular}{|l|l|l|l|l|l|}
\hline
$t_1t_2bcd$ & $t_1t_2bcd$ & $t_1t_2bcd$ & $t_1t_2bcd$ & $t_1t_2bcd$\\
\hline
\hline
00001 & 01010 & 11001 & 00100 & 11110\\
00010 & 01011 & 11010 & 10100 & 11111\\
01000 & 10000 & 11011 & 11100 & 12111\\
01001 & 11000 & 12011 & 11101 & 22111\\
\hline 
\end{tabular}
\caption{Positive topological charges of $D_5$.}\label{D5}
\end{table}

In Table \ref{D5} $t_1\equiv x_1+x_2$, $t_2\equiv z_1+z_2$.

Using relations

\begin{align}
t_1=h_2-h_3,\quad t_2=h_3-h_4,\quad b=h_1-h_2,\quad c=h_4+h_5,\quad d=h_4-h_5
\end{align}

Table \ref{D5} becomes Table \ref{D'5} which is obviously the table of positive roots of $SO(10)$.

\begin{table}
\begin{tabular}{|l|l|l|l|l|l|}
\hline
$h_1h_2h_3h_4h_5$ & $h_1h_2h_3h_4h_5$ & $h_1h_2h_3h_4h_5$ & $h_1h_2h_3h_4h_5$ & $h_1h_2h_3h_4h_5$\\
\hline
\hline
0001-1 & 00101 & 0100-1 & 1-1000 & 10001\\
00011 & 00110 & 01001 & 10-100 & 10010\\
001-10 & 01-100 & 01010 & 100-10 & 10100\\
0010-1& 010-10 & 01100 & 1000-1 & 11000\\
\hline 
\end{tabular}
\caption{Positive topological charges of $D_5$ in the new basis.}\label{D'5}
\end{table}

\subsection{$E_6$ quiver.}

After we set flux for the node X to zero the energy for bare monopoles is given by the expression

\begin{align}
& E=-(|s_1-s_2|+|l_1-l_2|+|l_1-l_3|+|l_2-l_3|+|m_1-m_2|+|p_1-p_2|)+\nonumber\\
& \frac12(|s_1|+|s_2|+|m_1-n|+|m_2-n|+|p_1-q|+|p_2-q|+|l_1-s_1|+|l_1-s_2|+|l_2-s_1|+|l_2-s_2|+\nonumber\\
& |l_3-s_1|+|l_3-s_2|+|l_1-m_1|+|l_1-m_2|+|l_2-m_1|+|l_2-m_2|+|l_3-m_1|+|l_3-m_2|+\nonumber\\
& |l_1-p_1|+|l_1-p_2|+ |l_2-p_1|+|l_2-p_2|+|l_3-p_1|+|l_3-p_2|)
\end{align}

Denote the six topological charges by 

\begin{align}
& k_1=n,\quad k_2=m_1+m_2,\quad k_3=l_1+l_2+l_3,\nonumber\\
& k_4=p_1+p_2,\quad k_5=q,\quad k_6=s_1+s_2.
\end{align}

The 36 topological scalars with positive values of topological charges reproduce 36 positive roots of $E_6$ (see \cite{Cornwell}). The rest 36 scalars have opposite topological charges appropriate for 36 negative roots.
\begin{table}
\begin{tabular}{|l|l|l|l|l|l|l|}
\hline
$k_1k_2k_3k_4k_5k_6$ &  $k_1k_2k_3k_4k_5k_6$ & $k_1k_2k_3k_4k_5k_6$ &  $k_1k_2k_3k_4k_5k_6$ & $k_1k_2k_3k_4k_5k_6$ &  $k_1k_2k_3k_4k_5k_6$\\
\hline
\hline
000001 &  001100  &  011001  &   012111  &  111100 & 112211\\
000010 &  001101  &  011100  &   012211  &  111101 & 122101\\
000100 &  001110  &  011101  &   100000  &  111110 & 122111\\
000110 &  001111  &  011110  &   110000  &  111111 & 122211\\
001000 &  010000  &  011111  &   111000  &  112101 & 123211\\
001001 &  011000  &  012101  &   111001  &  112111 & 123212\\
\hline 
\end{tabular}
\caption{Positive topological charges of $E_6$.}\label{E6}
\end{table}

\subsection{$E_7$ quiver.}

For the $E_7$ quiver theory we get 126 bare monopoles with energy $E=1$. Expressing the topological charges $k_i$ through magnetic charges

\begin{align}
k_1=d_1+d_2,\quad k_2=f_1+f_2+f_3,\quad k_3=g_1+g_2+g_3+g_4,\nonumber\\
k_4=h_1+h_2+h_3,\quad k_5=x_1+x_2,\quad k_6=b,\quad k_7=c_1+c_2,
\end{align}

we obtain exactly 126 roots of the Lie algebra $E_7$ as can be checked by comparing the spectrum of topological charges Table \ref{E7} with 63 positive roots of $E_7$ written down in \cite{Cornwell}.

\begin{table}
\begin{tabular}{|l|l|l|l|l|l|l|}
\hline
$k_i$ &  $k_i$ & $k_i$ &  $k_i$ & $k_i$ &  $k_i$\\
\hline
\hline
0000001 & 0011100 & 0111110 & 1110001 & 1122111 & 1232112\\                     
0000010 & 0011101 & 0111111 & 1111000 & 1122211 & 1232211\\                 
0000100 & 0011110 & 0121001 & 1111001 & 1221001 & 1232212\\         
0000110 & 0011111 & 0121101 & 1111100 & 1221101 & 1233211\\                
0001000 & 0100000 & 0121111 & 1111101 & 1221111 & 1233212\\             
0001100 & 0110000 & 0122101 & 1111110 & 1222101 & 1243212\\             
0001110 & 0110001 & 0122111 & 1111111 & 1222111 & 1343212\\             
0010000 & 0111000 & 0122211 & 1121001 & 1222211 & 2343212\\                              
0010001 & 0111001 & 1000000 & 1121101 & 1232101 & \\                               
0011000 & 0111100 & 1100000 & 1121111 & 1232102 & \\                                          
0011001 & 0111101 & 1110000 & 1122101 & 1232111 & \\
\hline 
\end{tabular}
\caption{Positive topological charges of $E_7$.}\label{E7}
\end{table}

\subsection{$E_8$ quiver.}

\begin{table}
\begin{tabular}{|l|l|l|l|l|l|l|}
\hline
$k_i$ &  $k_i$ & $k_i$ &  $k_i$ & $k_i$ &  $k_i$\\
\hline
\hline
00000001 & 00111111 & 01222211 & 11222101 &12322102  & 13432212\\
00000010 & 01000000 & 10000000 & 11222111 & 12322111 & 13433212\\
00000100 & 01100000 & 11000000 & 11222211 & 12322112 & 13433212\\
00000110 & 01100001 & 11100000 & 12210001 & 12322211 & 13443212\\
00001000 & 01110000 & 11100001 & 12211001 & 12322212 & 13543212\\
00001100 & 01111000 & 11110000 & 12211101 & 12332101 & 13543213\\
00001110 & 01111001 & 11110001 & 12211111 & 12332102 & 23432102\\
00010000 & 01111100 & 11111000 & 12221001 & 12332111 & 23432112\\
00011000 & 01111101 & 11111001 & 12221101 & 12332112 & 23432212\\
00011100 & 01111110 & 11111100 & 12221111 & 12332211 & 23433212\\    
00011110 & 01111111 & 11111101 & 12222101 & 12332212 & 23443212\\        
00100000 & 01210001 & 11111110 & 12222111 & 12333211 & 23543212\\
00100001 & 01211001 & 11111111 & 12222211 & 12333212 & 23543213\\   
00110000 & 01211101 & 11210001 & 12321001 & 12432102 & 24543212\\    
00110001 & 01211111 & 11211001 & 12321002 & 12432112 & 24543213\\          
00111000 & 01221001 & 11211101 & 12321101 & 12432212 & 24643213\\           
00111001 & 01221101 & 11211111 & 12321102 & 12433212 & 24653213\\              
00111100 & 01221111 & 11221001 & 12321111 & 12443212 & 24654213\\          
00111101 & 01222101 & 11221101 & 12321112 & 13432102 & 24654313\\          
00111110 & 01222111 & 11221111 & 12322101 & 13432112 & 24654323\\
\hline 
\end{tabular}
\caption{Positive topological charges of $E_8$.}\label{E8}
\end{table}

Finally, for the $E_8$ quiver theory the spectrum of topological charges
\begin{align}
& k_1=x_1+x_2,\quad k_2=c_1+c_2+c_3+c_4,\quad k_3=b_1+b_2+b_3+b_4+b_5+b_6,\nonumber\\
& k_4=a_1+a_2+a_3+a_4+a_5,\quad k_5=g_1+g_2+g_3+g_4,\quad k_6=f_1+f_2+f_3,\nonumber\\
& k_7=d_1+d_2,\quad k_8=h_1+h_2+h_3
\end{align}
on the energy level $E=1$ (Table \ref{E8}) coincides with 240 roots of $E_8$ the positive part of which can be compared with \cite{Cornwell}.

\bigskip
\bigskip

\section{Appendix B.}

\bigskip

\begin{center}
\begin{picture}(275,150)(5,-10)
\put(15,15){\circle{20}}
\put(13,13){1}
\put(25,25){m}
\put(25,15){\line(1,0){40}}
\put(75,15){\circle{20}}
\put(73,13){2}
\put(85,25){n}
\put(85,15){\line(1,0){40}}
\put(135,15){\circle{20}}
\put(133,13){3}
\put(145,25){l}
\put(135,25){\line(0,1){40}}
\put(135,75){\circle{20}}
\put(133,73){2}
\put(145,85){s}
\put(135,85){\line(0,1){40}}
\put(135,135){\circle{20}}
\put(133,133){1}
\put(145,15){\line(1,0){40}}
\put(195,15){\circle{20}}
\put(193,13){2}
\put(205,25){p}
\put(205,15){\line(1,0){40}}
\put(255,15){\circle{20}}
\put(253,13){1}
\put(265,25){q}
\put(113,-10){\small Figure B1. $E_6$ quiver.}
\end{picture}
\end{center}

\bigskip

\begin{center}
\begin{picture}(380,100)(-55,-10)
\put(-35,15){\line(1,0){40}}
\put(-45,15){\circle{20}}
\put(-48,13){1}
\put(15,15){\circle{20}}
\put(13,13){2}
\put(25,25){d}
\put(25,15){\line(1,0){40}}
\put(75,15){\circle{20}}
\put(73,13){3}
\put(85,25){f}
\put(85,15){\line(1,0){40}}
\put(135,15){\circle{20}}
\put(133,13){4}
\put(145,25){g}
\put(135,25){\line(0,1){40}}
\put(135,75){\circle{20}}
\put(133,73){2}
\put(145,85){c}
\put(145,15){\line(1,0){40}}
\put(195,15){\circle{20}}
\put(193,13){3}
\put(205,25){h}
\put(205,15){\line(1,0){40}}
\put(255,15){\circle{20}}
\put(253,13){2}
\put(265,25){x}
\put(265,15){\line(1,0){40}}
\put(315,15){\circle{20}}
\put(313,13){1}
\put(325,25){b}
\put(80,-10){\small Figure B2. $E_7$ quiver.}
\end{picture}
\end{center}

\bigskip

\begin{center}
\begin{picture}(440,100)(-115,-10)
\put(-95,15){\line(1,0){40}}
\put(-105,15){\circle{20}}
\put(-108,13){1}
\put(-35,15){\line(1,0){40}}
\put(-45,15){\circle{20}}
\put(-48,13){2}
\put(-35,25){d}
\put(15,15){\circle{20}}
\put(13,13){3}
\put(25,25){f}
\put(25,15){\line(1,0){40}}
\put(75,15){\circle{20}}
\put(73,13){4}
\put(85,25){g}
\put(85,15){\line(1,0){40}}
\put(135,15){\circle{20}}
\put(133,13){5}
\put(145,25){a}
\put(195,25){\line(0,1){40}}
\put(195,75){\circle{20}}
\put(193,73){3}
\put(205,85){h}
\put(145,15){\line(1,0){40}}
\put(195,15){\circle{20}}
\put(193,13){6}
\put(205,25){b}
\put(205,15){\line(1,0){40}}
\put(255,15){\circle{20}}
\put(253,13){4}
\put(265,25){c}
\put(265,15){\line(1,0){40}}
\put(315,15){\circle{20}}
\put(313,13){2}
\put(325,25){x}
\put(50,-10){\small Figure B3. $E_8$ quiver.}
\end{picture}
\end{center}

\bigskip

\begin{center}
\begin{picture}(270,280)(5,-10)
\put(15,15){\circle{20}}
\put(13,13){1}
\put(25,15){\line(1,0){40}}
\put(75,15){\circle{20}}
\put(73,13){3}
\put(85,15){\line(1,0){40}}
\put(135,15){\circle{20}}
\put(133,13){5}
\put(135,25){\line(0,1){40}}
\put(135,75){\circle{20}}
\put(133,73){4}
\put(135,85){\line(0,1){40}}
\put(135,135){\circle{20}}
\put(133,133){3}
\put(135,145){\line(0,1){40}}
\put(135,195){\circle{20}}
\put(133,193){2}
\put(135,205){\line(0,1){40}}
\put(135,255){\circle{20}}
\put(133,253){1}
\put(145,15){\line(1,0){40}}
\put(195,15){\circle{20}}
\put(193,13){3}
\put(205,15){\line(1,0){40}}
\put(255,15){\circle{20}}
\put(253,13){1}
\put(113,-10){\small Figure B4.}
\end{picture}
\end{center}

\section{Acknowledgements.}

I am grateful to Anton Kapustin for illuminating discussions.

\end{document}